\documentclass[aps,prd,reprint,preprintnumbers,floats,epsfig,nofootinbib,amssymb,nofootinbib]{revtex4}

\usepackage{slashed}
\usepackage{graphicx,color}
\usepackage{epsfig}
\usepackage{subfigure}
\usepackage{epsfig}
\usepackage{dcolumn}
\usepackage{bm}
\usepackage{color}
\usepackage{ulem}
\usepackage{enumitem}
\usepackage[colorlinks,
            linkcolor=black,
            anchorcolor=black,
            citecolor=black
            ]{hyperref}

\begin{document}

\baselineskip=15pt


\title{Structure Of Flavor Changing Goldstone Boson Interactions}

\author{Jin Sun$^1$\footnote{019072910096@sjtu.edu.cn}}
\author{Yu Cheng$^1$\footnote{chengyu@sjtu.edu.cn}}
\author{Xiao-Gang He$^{2,3}$\footnote{hexg@phys.ntu.edu.tw}}

\affiliation{${}^{1}$Tsung-Dao Lee Institute, and School of Physics and Astronomy, Shanghai Jiao Tong University, Shanghai 200240, China}
\affiliation{${}^{2}$Department of Physics, National Taiwan University, Taipei 10617, Taiwan}
\affiliation{${}^{3}$Physics Division, National Center for Theoretical Sciences, Hsinchu 30013, Taiwan}

\begin{abstract}
General flavor changing Goldstone boson (GB) interactions with fermions from a spontaneous global $U(1)_G$ symmetry breaking are discussed.
This GB may be the Axion, solving the strong QCD CP problem,
if there is a QCD anomaly for the assignments of quarks $U(1)_G$ charge. Or it may be the Majoron, producing seesaw Majorana neutrino masses by lepton number violation,
if the symmetry breaking scale is much higher than the electroweak scale. It may also, in principle, play the roles of Axion and Majoron simultaneously
as far as providing solution for the strong CP problem and generating a small Majorana neutrino masses are concerned.
Great attentions have been focused on flavor conserving GB interactions. Recently flavor changing Axion and Majoron models have been studied in the hope to find new physics from rare decays in the intensity frontier. In this work,
we will provide a systematic model building aspect study for flavor changing neutral current (FCNC) GB interactions in the fermion sectors, or separately in the quark, charged lepton and neutrino sectors and will identify in detail the sources of FCNC interactions in a class of beyond standard model with a spontaneous global $U(1)_G$ symmetry breaking.  We also provide a general proof of  the equivalence of using physical GB components and GB broken generators for calculating GB couplings to two gluons and two photons, and discuss some issues related to spontaneous CP violation models. Besides,
we will also provide some details for obtaining FCNC GB interactions in several popular models, such as the Type-I, -II, -III seesaw and Left-Right symmetric models, and point out some special features in these models.
\end{abstract}

\maketitle

\section{Introduction}

 A Goldstone boson (GB), a massless spin zero particle, from spontaneous symmetry break down of some global symmetries is an important result of quantum field theory~\cite{Nambu:1960tm,Goldstone:1961eq}. When the original symmetry is gauged, the GB would be ``eaten'' by gauge boson corresponding to the broken generator of the symmetry, so that it acquires the longitudinal component degrees of freedom. The Higgs mechanism~\cite{Englert:1964et,Higgs:1966ev,Guralnik:1964eu} for electroweak symmetry breaking and mass generation of  the standard model (SM) particles is a good example of this type. This mechanism has been verified experimentally by the discovery of the Higgs boson. If the original symmetry is a global symmetry, the GB will be a physical massless particle~\footnote{If there are anomalies at quantum level, the corresponding GB may gain a finite mass, such as QCD Axion~\cite{Weinberg:1977ma,Wilczek:1977pj} from Peccei-Quinn symmetry~\cite{Peccei:1977hh,Peccei:1977ur} breaking.}. When going beyond the SM there are well motivated theoretical models with additional broken symmetries leading to the existence of physical GB particles. Some of the interesting examples are the Axion~\cite{Weinberg:1977ma,Wilczek:1977pj} from Peccei-Quinn symmetry~\cite{Peccei:1977hh,Peccei:1977ur} breaking for solving the strong CP problem, and the Majoron~\cite{Chikashige:1980qk} from lepton number (LN) symmetry breaking for neutrino mass generation.

Goldstone bosons have many laboratory, astrophysical and cosmological implications~\cite{Cheng:1987gp,Kim:1986ax,DiLuzio:2020wdo,Ballesteros:2016euj}. However, no fundamental GB has been detected experimentally so far. New dedicated experiments have been/are being designed to detect physical effects of GB. There have been extensive studies in this area. A great attentions have been focused on flavor conserving GB interaction~\cite{Cheng:1987gp,Kim:1986ax,DiLuzio:2020wdo,Ballesteros:2016euj}.
Recently flavor changing axion models have  received more attentions in the hope to find new physics from rare decays in the intensity frontier. With several high luminosity facilities in running, such as the BESIII, LHCb, BELLE-II, in recent years, looking for GB at the intensity frontier has attracted a lot of attentions.  Flavor changing neutral current (FCNC)  induced by GB in rare decays is some of the promising places to look for signs of new physics beyond SM including effects of GB interactions.
There are some stringent constraints from data already~\cite{Celis:2014iua,Ema:2016ops,Heeck:2017wgr,Calibbi:2016hwq,
 Marciano:2016yhf,CidVidal:2018blh,Heeck:2019guh,Calibbi:2020jvd,MartinCamalich:2020dfe,Zyla:2020zbs,Cornella:2019uxs,Cheng:2020rla}.
 It has recently been shown that by measuring the polarization of final-state charged leptons, the chiral structure of FCNC GB interaction can also be studied~\cite{Cheng:2020rla}.

Some of the well motivated models having a GB are the Axion and Majoron models.  Many of the searches depend on how GB interacts with SM fermions. GB couplings to fermions not only have flavor conserving interactions, but also flavor changing ones.  This was known a long time ago with some interesting phenomena~\cite{Schechter:1981cv,Gelmini:1983ea,Anselm:1985bp,Berezhiani:1989fp,GonzalezGarcia:1988rw,Pilaftsis:1993af} and has attracted many attentions recently~\cite{Celis:2014iua,Heeck:2017wgr,Ema:2016ops,Heeck:2019guh,MartinCamalich:2020dfe}.

GB interaction with fermions is in derivative form and it is usually parameterized as the following
\begin{eqnarray}
L^c_{int} = {\partial_\mu a \over 2 f_a} \bar f_j \gamma^\mu (c^{jk}_V + c^{jk}_A\gamma_5) f_k\;,\label{Lint}
\end{eqnarray}
where $f$ stands for a quark  or a charged lepton or a light active neutrino, and $j\;,k$ are the generation indices. $f_a$ is the GB decay constant which sets the scale of  $U(1)_G$ symmetry breaking.  $c_{V,A}$ satisfy the condition $c^{\dagger}_{V/A} = c_{V/A}$ to have a hermitian interaction Lagrangian. The sizes of $c_{V,A}$ are model dependent.

 If neutrinos are Dirac particles, the GB interactions with neutrinos will be the same in form as given above. If neutrinos are Majorana particles, the form  will be modified.  Also if right-handed neutrinos $\nu_R$ are introduced to facilitate the seesaw mechanism, $\nu_L$ and $\nu^c_R$ will have different masses, this also modifies the form of the interaction.
The Lagrangian $L^\nu_{int}$  of GB interaction with neutrinos as appearing in seesaw models will have the following form
\begin{eqnarray}
L^\nu_{int} = {\partial_\mu a \over 2 f_a} \left (\bar \nu_{Lj} \gamma^\mu c^{jk}_{LL} \nu_{Lk}  + \bar \nu^c_{Rj} \gamma^\mu c^{jk}_{RR} \nu^c_{Rk}
+ (\bar \nu_{L j} \gamma^\mu c^{jk}_{LR} \nu^c_{Rk} + \mbox{H.c.})\right )\;.\label{Lintn}
\end{eqnarray}

The flavor changing GB interactions with fermions have a lot of interesting phenomena which can be used to discover a GB. These can be from rare decays of particles containing b, c and s quarks, $\tau$, $\mu$ charged lepton decays, neutrino decays, and B-, D-, K-meson and muonium oscillations, and also g-2 of charged leptons~\cite{Anselm:1985bp,Berezhiani:1989fp, Calibbi:2016hwq,CidVidal:2018blh,MartinCamalich:2020dfe,GonzalezGarcia:1988rw,Pilaftsis:1993af,Heeck:2017wgr,Heeck:2019guh,Calibbi:2020jvd,Marciano:2016yhf,Cornella:2019uxs,Cheng:2020rla}. In this work, we will not repeat to obtain the stringent constraints from various data, but to investigate some interesting features of FCNC GB interactions from a general $U(1)_G$ global symmetry break down beyond SM and some related issues.
This GB can be an Axion, a Majoron or a mixture of them, that is, a GB can play the role of the Axion and Majoron simultaneously~\cite{Mohapatra:1982tc} and has some interesting features~\cite{Ballesteros:2016euj}.
In a concrete model, there are usually other Higgs doublets besides the SM one. In general the additional Higgs may also mediate FCNC interactions~\cite{Glashow:1976nt}. These new Higgs bosons all have masses and some of them can be much larger than the electroweak scale. For a massless GB, its FCNC effects will be different. So we will concentrate on FCNC structure of a GB. Note here that FCNC processes can also be generated at loop level where the strength is suppressed. So in this paper we only consider the tree level interactions.

The paper is arranged in the following way. In section II, we provide a systematic model building aspect study for GB interactions in both the quark and lepton sectors with a simple way to identify GB components, and to obtain GB-fermion interactions. For neutrino sector, we take Type-I seesaw as the prototype of model to study. In section III, we discuss under what conditions the general GB can be viewed as the usual Axion or Majoron. We provide a general proof of the equivalence of using physical GB components and GB broken generators for calculating Axion couplings to two gluons and two photons. In section IV, we identify in details the sources for FCNC GB interactions, and discuss how spontaneous CP violation may affect GB-fermion interactions. In section V, we discuss some interesting features of GB interactions with fermions in Type-II, -III seesaw models and Left-Right symmetric models. In section VI, we provide our conclusions.

\section{A general global $U(1)_G$ model and its goldstone-fermion interactions}

In the standard model, with the SM gauge particles, the standard three generations of fermions and the Higgs boson doublet, and also with the fully allowed Yukawa couplings of the Higgs doublet with SM fermions, the model contains several accidental global symmetries, such as the lepton number $L$ and baryon number $B$.
Each of them can be identified with a global $U(1)$ symmetry respectively~\footnote{non-perturbative effects, such as instanton effects, will break $B+L$~\cite{tHooft:1976rip}.}. When going beyond the minimal SM, by adding new particles, the global lepton and baryon number symmetries can occur spontaneously broken to produce GBs. There are five types of fermions in SM, such as the three generations of left-handed quark doublets $Q^j_L$ or lepton doublets $L^j_L$ and the right-handed quark singlets $U^j_R$, $D^j_R$ or charged lepton singlets $E_R$.
If one switches off the Yukawa couplings, each type of fermions poses a $U(3)$ global symmetry so that the model has a $U(3)^5$ global symmetry. If further introducing right-handed neutrinos, the global symmetry can be even larger. 
Starting with such a theory at a high energy scale and then breaking these global symmetries spontaneously down to lower energies with only a $U(1)$ baryon and a $U(1)$ lepton numbers as the usual SM, it will result in many GBs associated with the broken generators.
Depending on the structure of the vacuum expectation values (vevs) of the new scalar particles in the model, the symmetry breaking chains may have a complicated route for having a phenomenologically acceptable model. The complicated analysis
may blur our aim to have a clear picture for the properties about GB itself and it has been beyond the scope of our paper. Therefore we will limit our discussions to the specific class of models, which only has an additional global $U(1)_G$ symmetry occurring spontaneously broken by vevs of some necessary new introduced scalar particles besides three generations of fermions, so that we can obtain detailed information that how this GB interacts with fermions to generate FCNC interactions.
This $U(1)_G$ can be the Peccei-Quinn symmetry for solving the strong CP problem or
lepton number (LN)  symmetry in connection with Majoron models or some other flavor symmetries, which depends on how the $U(1)_G$ acts on the particles in the model.

 In a general form, we assume fermions in the model transform under $U(1)_G$ as
\begin{eqnarray}
f^j_L \to e^{i X^j_L} f^j_L\;,\;\;\;\;f^j_R \to e^{i X^j_R} f^j_R\;,
\end{eqnarray}
$f_{L, R}$ are the fermions in the SM with $SU(3)_C\times SU(2)_L\times U(1)_Y$ gauge symmetries. For quarks, $f^j_L$ is $Q^j_L: (3,2,1/6)(X_L^{q j})$, $f^j_R$ is one of $U_R^{j}: (3,1,2/3)(X^{u j}_R)$, or $D_R^{j}: (3,1,-1/3)(X^{d j}_R)$, and
for leptons, $f^j_L$ is $L^j_L: (1,2,-1/2) (X^{l j}_L)$, $f_R^j$ can be $E_R^{j}: (1,1,-1)(X^{e j}_R)$. Since $X_L^{qj}$ and $X_L^{lj}$
contain $u_L^j,\; d^j_L$ and $\nu_L^j,\; e^j_L$, we indicate their individual $U(1)_G$ charges as $X^{uj}_L = X^{dj}_L = X^{qj}_L$ and $X_L^{\nu j} = X_L^{ej} = X_L^{lj}$ for conveniences. If there are right-handed neutrinos,
$f^j_R$ is $\nu_R^j: (1,1,0)(X^{\nu j}_R)$. The quantum numbers in the brackets correspond to $SU(3)_C$, $SU(2)_L$, $U(1)_Y$ and $U(1)_G$, respectively. The diagonal matrix diag$(X^{f 1}_{L,R}, X^{f 2}_{L,R}, X^{f 3}_{L,R})$ in flavor space  will be indicated by a diagonal matrix $X_{L,R}^f$.
In general there are several Higgs doublets $H^{u,d,e,\nu}_{jk}$ transforming as $(1,2,1/2) (X^{q,l \; j}_L - X^{u,d,e,\nu\;k}_R)$ which couple to fermions,
\begin{eqnarray}
L_Y = - \bar Q_L^j Y^{jk}_u \tilde H^u_{jk} U_R^k - \bar Q_L^j Y^{jk}_d H^d_{jk} D_R^k - \bar L_L^j Y^{jk}_e H^e_{jk} E_R^k
- \bar L_L^j Y^{jk}_\nu \tilde H^\nu_{jk} \nu_R^k + \mbox{H.c.}\;.
\end{eqnarray}
 In the above $j$ and $k$ are summed over generation indices. The superscripts (subscripts) $u$, $d$, $e$ and $\nu$ on Higgs doublets (Yukawa couplings) are summed over Higgs doublets in the model.
In component form
\begin{eqnarray}
H^a_{jk} = \left ( \begin{array}{cc}
h^{a+}_{jk}\\
\\
{1\over \sqrt{2}}(v^a_{jk} + h^a_{jk} + i I_{jk}^a)
\end{array}
\right )\;.
\end{eqnarray}
When the Higgs bosons develop vevs, $v^{u,d,e,\nu} _{jk}$, the electroweak symmetry $SU(2)_L \times U(1)_Y$
is broken down to electromagnetic symmetry $U(1)_{em}$, and at the same time the $U(1)_G$ is also broken. Non-zero vevs will give the masses of fermions and gauge bosons $W$, $Z$.

If at the same time the singlets $S_{jk}$ are introduced with $U(1)_G$ charge $-(X^{\nu j}_R +X^{\nu k}_R)$, one can also have the
terms $- (1/2) \bar \nu^{c j}_R Y^{jk}_s S_{jk}\nu^k_R$. Here the superscript c indicates the charge conjugated field. If there are more than one singlet,  $Y^{jk}_s S_{jk}$ implies summation of singlets contributions.
$S_{jk} = (1/\sqrt{2})(v^s_{jk}+ R^s_{jk} + i I^s_{jk})$. When the vevs of $v^s_{jk}/\sqrt{2}$ become non-zero and are larger than $v^{u,d,e,\nu}_{jk}$, the Type-I seesaw~\cite{Minkowski:1977sc,Yanagida:1980xy,type1-seesaw,Glashow:1979nm,Mohapatra:1979ia,Schechter:1981cv} mechanism  will be in effective to provide small Majorana masses for light neutrinos. The singlets can also play the role of making possible dangerous GB interactions invisible as in the DFSZ invisible Axion model~\cite{Dine:1981rt,Zhitnitsky:1980tq}.

 One may wonder whether one just needs to consider the effects where only one global $U(1)_G$ symmetry in addition to the SM gauge is broken spontaneously when the model has the above complicated scalar particle contents.
This replies on how model dependent new scalars are introduced. Since the singlets can have arbitrary $U(1)_G$ charges, one can choose appropriate charges for the singlets so that in the model only one global $U(1)_G$ symmetry is broken spontaneously. Several example models of this type with reasonably complicated Higgs structure have been discussed in Ref.~\cite{Sun:2020iim}. One may also resort to higher dimensional operators to  break appropriately unnecessary left-over symmetries~\cite{Ema:2016ops,Calibbi:2016hwq} except the $U(1)_G$ at the beginning of the symmetry breaking. Our discussions in the following apply to this class of  renormalizable models.

As mentioned before, the non-zero vevs of scalars $H^a_{jk}$ and $S_{jk}$ not only break  the electroweak symmetry to provide the longitudinal components of weak gauge bosons $W$ and $Z$, but also break the global $U(1)_G$ symmetry to result in a massless GB.
The vector $z$ ``eaten'' by $Z$ boson, in the basis $\vec I = (I^u_{jk}, I^d_{jk},I^e_{jk},I^\nu_{jk},I^s_{jk})$, is
given by
\begin{eqnarray}
\vec z = (v^u_{jk},\; v^d_{jk},\; v^e_{jk},\; v^\nu_{jk},\; 0)\;,
\end{eqnarray}
and the $U(1)_G$ broken generator vector $A$ is given by
\begin{eqnarray}
\vec A = \left (-(X^{u j}_L - X^{u k}_R)v^u_{jk},\; (X^{d j}_L - X^{d k}_R)v^d_{jk},\; (X^{e j}_L - X^{ e k}_R)v^e_{jk},\; -(X^{\nu j}_L - X^{\nu k}_R )v^\nu_{jk},\; -(X^{\nu j}_R +X^{\nu k}_R)v^s_{jk} \right )\;.
\end{eqnarray}

The physical GB  in this model should be the linear combination $a = \vec a \cdot \vec I^T$, which is orthogonal to $z = \vec z\cdot \vec I^T$. The corresponding vector form is $\vec a = \alpha \vec z + \vec A$. The requirement that $\vec a\cdot \vec z^T = 0$ dictates
$\alpha \sim {-\vec A\cdot \vec z^T/\vec z \cdot \vec z^T}$. Therefore $\vec a$ is given by~\cite{Sun:2020iim}
\begin{eqnarray}
\vec a = {1\over N_\alpha} (\bar v^2 \vec z - v^2 \vec A)\;,
\end{eqnarray}
where $N_\alpha$ is a normalization constant to ensure $\vec a\cdot \vec a^T=1$, and
\begin{eqnarray}
&&v^2 = \vec z\cdot \vec z^T  = (v^u_{jk})^2 + (v^d_{jk})^2+(v^e_{jk})^2+(v^\nu_{jk})^2\;,\nonumber\\
&&\bar v^2 =\vec A\cdot \vec z^T = -(X^{u j}_L - X^{u k}_R)(v^u_{jk})^2+ (X^{d j}_L - X^{d k}_R)(v^d_{jk})^2+(X^{e j}_L - X^{ e k}_R)(v^e_{jk})^2-(X^{\nu j}_L - X^{\nu k}_R )(v^\nu_{jk})^2\;.
\end{eqnarray}
Expressing the physical GB, $a = \vec a\cdot \vec I^T$,  in terms of $I^a_{jk}$ , we have
\begin{eqnarray}
a = {1\over N_\alpha}  \left [ \left ((X^{p l}_L - X^{p m}_R)   -  (X^{qj}_{L} - X^{qk}_R)\right ) (v^p_{lm})^2 v^{q}_{jk}sign(q)I^q_{jk} + (X^{\nu j}_R +X^{\nu k}_R) (v^p_{lm} )^2 v^s_{jk} I^s_{jk}\right ]\;. \label{axion-field}
\end{eqnarray}
In the above, $j,\;k$, and $l,\;m$ are summed over flavor spaces in each sector, and p, q are summed over $u,\;d,\;e,\nu$. Here sign(q) takes ``$-$'' for
$q=u,\;\nu$ and ``$+$'' for $q=d,e$.
 The GB field is uniquely determined. We comment that if there are other additional global symmetry breaking spontaneously,  it would involve in identifying the other associated GB fields~\cite{Berezhiani:1989fp}. The above results will not apply again.

The above shows that $I^q_{jk}$ and $I^s_{jk}$ contain the GB $a$ with amplitude $(1/N_\alpha)((X^{p l}_L - X^{p m}_R) - (X^{qj}_{L} - X^{qk}_R) ) (v^p_{lm})^2 v^{q}_{jk}sign(q)$ and   $(1/N_\alpha)(X^{\nu j}_R + X^{\nu k}_R)(v^p_{lm})^2 v^{s}_{jk}$, respectively. The Yukawa couplings of GB $a$ to fermions along with the mass terms are given by
\begin{eqnarray}
L_Y &=& - \bar U^j_L M^{jk}_u \left [ 1 + i a {v^2\over N_\alpha} \left (-{\bar v^2\over v^2}  -  (X^{uj}_{L} - X^{uk}_R)\right ) \right ] U^k_R - \bar D^j_L M^{jk}_d \left [ 1 + i a {v^2\over N_\alpha} \left ({\bar v^2\over v^2}  -  (X^{d j}_{L} - X^{d k}_R) \right )\right ] D^k_R \nonumber\\
&&- \bar E^j_L M^{jk}_e \left  [ 1 + i a {v^2\over N_\alpha}\left  ({\bar v^2\over v^2}  -  (X^{ej}_{L} - X^{ek}_R)\right ) \right ] E^k_R - \bar \nu^j_L M^{jk}_D\left  [ 1 + i a {v^2\over N_\alpha} \left (- {\bar v^2\over v^2}   -  (X^{\nu j}_{L} - X^{\nu k}_R) \right ) \right ]\nu^k_R \nonumber\\
&&- {1\over 2}  \bar \nu^{c j}_R M^{jk}_R \left  (1 + i a {v^2\over N_\alpha} (X^{\nu j}_R + X^{\nu k}_R) \right )\nu^k_R  + \mbox{H.c.}\;,
\end{eqnarray}
where $M^{jk}_q$ are mass matrices for up quark $M_u$, down quark $M_d$, charged lepton $M_e$ and neutrino $M_\nu$. They are given by\begin{eqnarray}
&&M^{jk}_u =  {Y^{jk}_u v^u_{jk}\over \sqrt{2}}\;,\hspace{1.1cm}M^{jk}_d =  {Y^{jk}_d v^d_{jk}\over \sqrt{2}}\;,\hspace{1.4cm} M^{jk}_e =  {Y^{jk}_e v^e_{jk}\over \sqrt{2}}\;,\nonumber\\
&&M^{jk}_\nu = \left (\begin{array}{ll}
0&\;\;M_D\\
M^T_D&\;\;M_R
\end{array} \right )^{jk}\;,\;\;\mbox{with}\;\; M^{jk}_D =  {Y^{jk}_\nu v^\nu_{jk}\over \sqrt{2}}\;,\;\;\;\;M^{jk}_R =  {Y^{jk}_s v^s_{jk}\over \sqrt{2}}\;.
\end{eqnarray}
The above mass matrices should be summed over contributions from different pieces of each vev $v^q_{jk}$ for each ``$q$'' type of fermions. Note that here $j$ and $k$ are not summed.

From the above Yukawa couplings, we can identify the fermion current interacting with derivative form of $a$,  $L_Y  \to L_{af} = \partial_\mu a  j^\mu_{af}$, with the help of the equations of motion as
\begin{eqnarray}\label{jaf} \label{mast-eq}
j^\mu_{af}=
&& {\bar v^2\over 2 N_\alpha} \left ( ( \bar U_L \gamma^\mu  U_L -  \bar U_R \gamma^\mu  U_R)  -  ( \bar D_L\gamma^\mu D_L - \bar D_R\gamma^\mu D_R) -  (\bar E_L\gamma^\mu E_L - \bar E_R\gamma^\mu E_R )+  2 \bar \nu_L \gamma^\mu \nu_L \right )
\nonumber\\
&&+ {v^2\over  N_\alpha} (\bar U_L  X^u_L \gamma^\mu U_L + \bar U_R  X^u_R   \gamma^\mu U_R) +  {v^2\over N_\alpha} (\bar D_L X^d_L \gamma^\mu D_L + \bar D_R  X^d_R \gamma^\mu D_R )\\
&& +   {v^2\over N_\alpha} (\bar E_L X^e_L \gamma^\mu E_L + \bar E_R  X^e_R \gamma^\mu E_R)+ {v^2\over N_\alpha} \left ( \bar \nu_L X^\nu_L \gamma^\mu \nu_L + \bar \nu_R X^\nu_R \gamma^\mu \nu_R \right )\;.\nonumber
\end{eqnarray}
We identify $1/f_a = v^2/N_\alpha$.

Note that the following relation holds,
\begin{eqnarray}
\bar \nu_L X^\nu_L \gamma^\mu \nu_L + \bar \nu_R X^\nu_R \gamma^\mu \nu_R = (\bar \nu_L, \bar \nu^c_R) X^\nu\gamma^\mu \left ( \begin{array}{l}
\nu_L\\ \nu^c_R \end{array}
\right),
\end{eqnarray}
where $X^\nu$ is a $6\times 6$ diagonal matrix with non-zero entries to be $(X^\nu_L, -X^\nu_R) = (X^{\nu 1}_L, X^{\nu 2}_L, X^{\nu 3}_L, - X^{\nu 1}_R, - X^{\nu 2}_R, -X^{\nu 3}_R)$.

 The above discussions can be easily extended to models having different scalars and fermions with different $SU(2)_L$ representations, such as the Type-II and Type-III seesaw models with FCNC GB interactions to be discussed later.
The same method can be applied to construct GB components in those models. In the models we discussed, FCNC GB interactions are generated at the tree level. One can also generate FCNC GB interactions at loop levels~\cite{Heeck:2019guh}. However, this will not be our aim in this paper and so it will not be discussed further.

Early invisible Axion models~\cite{Zhitnitsky:1980tq,Dine:1981rt} and many of variants without addressing neutrino masses can be obtained by dropping the last term in Eq.~(\ref{mast-eq}).
In most of the models, each type of the three generations of fermions is assigned to same $U(1)_G$ so that no FCNC GB interactions can be generated. Recently people paid more attention  to the models with different $U(1)_G$ charges for different generations of the type covered by Eq.~(\ref{mast-eq}) to generate FCNC GB interactions. Some of these models are discussed in Refs.~\cite{Celis:2014iua,MartinCamalich:2020dfe}.

The original Majoron model in Ref.~\cite{Chikashige:1980qk} is obtained by just introducing one Higgs doublet with a zero global lepton number, a singlet and leptons with all the same lepton number. Soon after it was realized that in such a model there are FCNC GB interactions and also some more elaborated models were constructed~\cite{Schechter:1981cv,Gelmini:1983ea,GonzalezGarcia:1988rw,Berezhiani:1989fp,Heeck:2019guh}.

A GB may play the role of both Axion and Majoron~\cite{Mohapatra:1982tc}. In fact  the models taking into account neutrino mass generations and also GB interactions, usually mix the role of Axion and Majoron. Most of models can be obtained  by assigning different $U(1)_G$ charges \cite{DiLuzio:2020wdo,Calibbi:2020jvd} to different generations of fermion or by adding terms like the last term in Eq.~(\ref{mast-eq}) for neutrino masses to generate FCNC GB interactions~\cite{Sun:2020iim,Cheng:2020rla,Chen:2007nx,He:2010hz,Pan:2020qqd}.

Our discussions so far do not include models with additional gauge symmetries. But this can be easily implemented by focusing on what symmetries are broken by the scalar vevs and reading off the GB components using the method described earlier. Along with more gauge symmetries are broken by the scalar vevs, the analysis becomes more complicated~\cite{Mohapatra:1982tc,Grimus:1982qu}, but the way of identifying the GB discussed earlier still applies. To make the points more explicit, we will provide some illustrations for Left-Right symmetric model~\cite{Mohapatra:1974gc,Senjanovic:1975rk} later.

There are also models with non-renormalizable GB couplings~~\cite{Ema:2016ops,Calibbi:2016hwq}. Our method still can  be easily extended to this type of models since the identification of GB components for each of the scalar boson with vev breaking the symmetries is the same as discussed before.
But allowing non-renormalizable terms in the model provides another type of source of FCNC GB interactions. An example of this type of models is the flaxion model discussed in Ref.~\cite{Ema:2016ops,Calibbi:2016hwq}. In this model besides the SM particles, a singlet S with non-trivial $U(1)_G$ charge is introduced so that one adds additional terms of the type $y_{j k}^{f}\left(S/M\right)^{n_{kj}^{f}} \overline{f_L}_{j} H f_{R k}$.
The Higgs doublet does not have $U(1)_G$ charge. The $U(1)_G$ charge is balanced by the fermion $f_{L,R}$ $U(1)_G$ charges. The vev of the singlet $S$ does not break SM symmetry, but provide the only source for $U(1)_G$ breaking.  The imaginary  of $S$, $a$, is the GB in the model. Expanding additional Yukawa  couplings around the vacuum, the GB coupling to fermions becomes $i M_{jk} n_{jk}$ which are in general not simultaneously diagonalized depending on the choice of $n_{jk}$ and therefore the FCNC GB interaction can arise.
This class of models have simple scalar sector at the expenses of models with non-renormalizable interactions. We consider renormalizable models more attractive and therefore will work with this class of models.

\section{Goldstone boson as Axion or Majoron}

As mentioned before, the GB may or may not be a usual Axion or  Majoron. Here we make a rough distinction among them depending on their primary role in addressing some physics problems.
The massless GB will become massive if  the relevant $U(1)_G$ charge assignments have $SU(3)_C$ anomalies, then this model can be used to solve the strong CP problem. The GB in such models can be viewed as an Axion and the $U(1)_G$ can be identified as a variant of  the $U(1)_{PQ}$. The condition is to have
\begin{eqnarray}
Tr(X^u_R - X^u_L )+Tr(X^d_R - X^d_L ) \neq 0\;.
\end{eqnarray}
This can be understood from a possible GB-gluon coupling $ a \tilde G^{a \mu\nu} G^a_{\mu\nu}$ by calculating the triangle diagram using the current in Eq.~(\ref{jaf}). We have~\cite{Cheng:1987gp,Kim:1986ax,DiLuzio:2020wdo}
\begin{eqnarray}
L_{ag} = a {g^2_3\over 16 \pi^2} N(X)T(q) \tilde G^{a \mu\nu} G^a_{\mu\nu} = {\alpha_s\over 8 \pi} {a\over f_a} \tilde G^{a \mu\nu} G^a_{\mu\nu}\;,
\end{eqnarray}
where $g_3$ is the $SU(3)_C$ gauge coupling constant, and $T(q) $ is the generator of $SU(3)_C$ for color triplet quarks defined by $Tr (T^a T^b) = T(q) \delta^{ab}=\delta^{ab}/2$. $N(X) = N^u(X) + N^d(X)$. Here the superscripts indicate the contributions from up- and down-type quarks running in the loop of the triangle diagram. They are given by
\begin{eqnarray}
N^u(X) = N_G{\bar v^2\over N_\alpha} + {v^2\over N_\alpha} Tr(X^u_R - X^u_L )\;,\;\;N^d(X) = - N_G {\bar v^2\over N_\alpha} + {v^2\over N_\alpha} Tr(X^d_R - X^d_L )\;.
\end{eqnarray}
As long as $N(X) = (v^2/ N_\alpha) Tr(X^u_R - X^u_L + X^d_R - X^d_L)$ is not zero, there is a color anomaly. This makes the GB to be massive and play the role of the usual Axion\cite{Mohapatra:1982tc,Ballesteros:2016euj}.

Here we would like to make a comment on the relation of $a$ couplings to two gluons and two photons.
GB coupling to two photons of the type $a \tilde F^{\mu\nu} F_{\mu\nu}$ will be generated by just replacing gluons by photons in the above mentioned triangle diagram. We have
\begin{eqnarray}
L_{a\gamma} = a {e^2 \over 16 \pi^2} \tilde E(X) \tilde F^{ \mu\nu} F_{\mu\nu}\;,\label{Lagamma}
\end{eqnarray}
where $\tilde E(X) =E^u(X) + E^d(X) + E^e(X)$. The superscripts indicate the contributions from quarks and charged leptons running in the loop. They are given by
\begin{eqnarray}
&&E^u(X) = Q^2_u N^q_c N^u(X)\;,\;\;E^d(X) = Q^2_d N^q_c N^d(X)\;,\nonumber\\
&&E^e(X) = Q^2_e N_c^e \left (- N_G {\bar v^2\over N_\alpha} + {v^2\over N_\alpha} Tr(X^e_R - X^e_L )\right )\;.
\end{eqnarray}
Here $N^{q}_c = 3$ and $N^e_c = 1$ are the effective number of color for quarks and charged leptons, respectively. The above method is referred as calculation using the physical GB.

In the literature for Axion models, the GB-two-photon coupling is usually written as~\cite{DiLuzio:2020wdo}
\begin{eqnarray}
L_{a\gamma} = a {e^2 \over 16 \pi^2}  E(X) \tilde F^{\mu\nu} F_{\mu\nu}= {1\over 4} a g^0_{a\gamma} \tilde F^{\mu\nu} F_{\mu\nu}\;,\label{Lgamma}
\end{eqnarray}
Where $g^0_{a\gamma} = (\alpha_{em}/2\pi f_a) E(X)/N(X)$ with $E(X) = {v^2\over N_\alpha} Tr((X^u_R - X^u_L)Q^2_u N^q_c +  (X^d_R - X^d_L)Q^2_d N^q_c+ (X^e_R - X^e_L)Q^2_e N^e_c)$. This method is referred as calculation using the broken generators.

The above gives the same result as Eq.~(\ref{Lagamma}),  if $(Q^2_u - Q^2_d)N^q_c - Q^2_e N^e_c =0$, that is $E(X) = \tilde E(X)$. This condition is actually one of the gauge anomaly free  conditions~\cite{Bouchiat:1972iq,Geng:1988pr}
\begin{eqnarray}
I_3^u Q^2_uN^u_c + I^d_3 Q^2_dN^d_c + I_3^e Q^2_e N^e_c =0\;.
\end{eqnarray}
Here $I^f_3$ is the value of the third weak isospin component of the ``$f$th'' fermion.
Therefore this condition is guaranteed for a gauge anomaly free theory to eliminate the term proportional to $\bar v^2$ related to the component ``eaten'' by Z-boson,  which results in the same results obtained as using broken PQ generator. The above provides a general proof as discussed in Ref.~\cite{Sun:2020iim}. The results are completely fixed by the $U(1)_G$ charges $X^f_{L,R}$ and the kind of colored, and charged particles in the model.

Note that if there is no color anomaly for $U(1)_G$, that is $N(X) = 0$ as  in the Majoron models in Ref.~\cite{Chikashige:1980qk, Cheng:2020rla}, the situation will be different. In this case to avoid that $N(X)$ appears in the denominator of $g^0_{a\gamma} = (\alpha_{em}/2\pi f_a) E(X)/N(X)$, it is better to use $g^0_{a\gamma} = (\alpha_{em} / 4 \pi) E(X)$ directly.

Majoron is also another commonly studied GB which results from spontaneous break down of lepton number, like in the Type-I seesaw model~\cite{Chikashige:1980qk}. Therefore there is no color anomaly for GB produced by lepton number breaking.

From our discussion in previous section, the GB can in general have color anomaly and also break lepton number, therefore the GB can be viewed as an Axion and Majoron simultaneously~\cite{Mohapatra:1982tc}. The GB also exists  other names~\cite{Ma:2017vdv}, such as Familon~\cite{Anselm:1985bp,Berezhiani:1989fp}, and Arion~\cite{Anselm:1982ip}, which can be considered as special cases discussed here. But whichever name the GB has, it results from a global $U(1)$ symmetry breaking.

\section{Flavor changing Goldstone boson interactions}

We now discuss how FCNC GB interactions with fermions emerge. The relevant information is contained in the GB current in Eq.~(\ref{jaf}). The flavor changing nature of the interaction can be easily seen in the mass eigen-state basis.
The mass matrices for fermions can be diagonalized by bi-unitary transformation to the diagonal ones, $\hat M_f = V^f_L M_f V^{f\dagger}_R$. In the mass-eigen basis, the GB interaction current $j^\mu_{ac}$ with   quarks and charged leptons is given by
\begin{eqnarray}
j^\mu_{ac} =
 &&- {\bar v^2\over 2 N_\alpha}  (\bar U^m \gamma^\mu \gamma_5 U^m -  \bar D^m\gamma^\mu \gamma_5 D^m -  \bar E^m\gamma^\mu \gamma_5 E^m)
+ {v^2\over N_\alpha} (\bar U^m_L V^u_L X^u_LV_L^{u \dagger} \gamma^\mu U^m_L + \bar U^m_R V^{u}_R X^u_R V^{u \dagger}_R \gamma^\mu U^m_R) \nonumber\\
&&+  {v^2\over N_\alpha} (\bar D^m_L V^d_LX^d_LV^{d\dagger}_L \gamma^\mu D^m_L + \bar D^m_R V^{d}_R X^d_RV^{d \dagger}_R \gamma^\mu D^m_R )
 +   {v^2\over N_\alpha} (\bar E^m_L V^e_L X^e_L V^{e\dagger}_L \gamma^\mu E^m_L + \bar E^m_R V^{e}_R X^e_R V^{e \dagger}_R \gamma^\mu E^m_R)
\;. \label{jac}
\end{eqnarray}
Here $X^f_{L,R}$ are diagonal matrices with the diagonal entries given by $(X^{f 1}_{L,R}, X^{f 2}_{L,R},X^{f 3}_{L,R})$. $f^m$ indicates the mass eigen-states. We will drop the superscript ``$m$'' to keep notation simple unless stated otherwise.
It is clear that when $X^f_{L,R}$ are not proportional to unit matrix the GB current is not diagonal in the mass eigen-state basis and therefore flavor changing interaction emerges.

The GB decay constant $f_a$ is identified by the relation $1/f_a = v^2/N_a$.
The off-diagonal elements for $c_V$ and $c_A$ in Eq.~(\ref{Lint}) are given by $(V^f_R X^f_R V^{f\dagger}_R + V^f_L X^f_L V^{f\dagger}_L)$ and $(V^f_R X^f_R V^{f\dagger}_R - V^f_L X^f_L V^{f\dagger}_L)$. For the diagonal elements, $\pm(\bar v^2/v^2)$ needs to be added to $c^{jk}_A$ entries with ``-'' for up-quarks, and ``+'' for down-quarks and charged leptons.  If $X_{L, R}^f$ entries are  order $O(1)$ and  have no accidental cancellations, $c_{V, A}$ can be order $O(1)$.

Similarly, GB couplings to neutrinos can be worked with some modifications. We provide some details here. The mass matrix $M_\nu$ for neutrinos is diagonalized by a $6\times 6$ unitary matrix $\hat M_\nu = V^\nu_L M_\nu V^{\nu T}_L$.
Writing $V_L^\nu$ into $3\times 3$ matrices blocks, we have
\begin{eqnarray}
V^\nu_L = \left ( \begin{array}{ll}
V^\nu_{LL}\;\;&V^\nu_{LR}\\
\\
V^\nu_{RL}\;\;&V^\nu_{RR}
\end{array}
\right )\;,
\end{eqnarray}
we have the current $j^\mu_{a\nu}$ for neutrinos given by
\begin{eqnarray}
j^\mu_{a\nu} =
 &&   {\bar v^2\over N_\alpha} (\bar \nu_LV^\nu_{LL} + \bar \nu^c_R V^\nu_{RL} )\gamma^\mu (V^{\nu\dagger}_{LL} \nu_L
+ V^{\nu \dagger}_{RL} \nu^c_R)\nonumber\\
&& + {v^2\over N_\alpha} \left ( (\bar \nu_LV^\nu_{LL} + \bar \nu^c_R V^\nu_{RL} ) X^\nu_L \gamma^\mu (V^{\nu\dagger}_{LL} \nu_L
+ V^{\nu \dagger}_{RL} \nu^c_R) + (\bar \nu^c_L V^{\nu *}_{LR} + \bar \nu_R  V^{\nu *}_{RR})X^\nu_R \gamma^\mu (V^{\nu T}_{LR} \nu_L ^c
+ V^{\nu T}_{RR} \nu_R) \right )
\;. \label{janu}
\end{eqnarray}
Again $f_a$ is identified by the relation $1/f_a = v^2/N_a$. Compared with Eq.~(\ref{Lintn}), we have
\begin{eqnarray}
&&c_{LL} = 2({\bar v^2\over v^2} V^\nu_{LL}V^{\nu \dagger}_{LL} + V^\nu_{LL}X^\nu_L V^{\nu \dagger}_{LL} - V^\nu_{LR}X^\nu_R V^{\nu \dagger}_{LR} )\;,\nonumber\\
&&c_{RR} =  2({\bar v^2\over v^2} V^\nu_{RL}V^{\nu \dagger}_{RL} + V^\nu_{RL}X^\nu_L V^{\nu \dagger}_{RL} - V^\nu_{RR}X^\nu_R V^{\nu \dagger}_{RR} )\;,\\
&&c_{LR} =  2({\bar v^2\over v^2} V^\nu_{LL}V^{\nu \dagger}_{RL} + V^\nu_{LL}X^\nu_L V^{\nu \dagger}_{RL} - V^\nu_{LR}X^\nu_R V^{\nu \dagger}_{RR} )\;.\nonumber
\end{eqnarray}

From the above, we see that there are more possibilities that FCNC interaction can emerge due to seesaw mass matrix diagonalization.
For example, as $V^\nu_{LL}$ are not unitary in general, FCNC interaction exists  in $a \bar \nu_L\nu_L$ interaction with amplitude proportional to $V^{\nu}_{LL} V^{\nu\dagger}_{LL}$. Since $V^\nu_{LL}$ should be close to the unitary $V_{PMNS}$ matrix, the FCNC interaction is naturally small.  The FCNC interaction can also occur, similar to the quarks and charged leptons if $X^{\nu }_{L,R}$ are not proportional to unit matrix. Even, $X^\nu_L$ and $X^\nu_R$ are separately proportional to unit matrix, FCNC interactions can still occur if the $6\times 6$ diagonal matrix $X^\nu$ is not proportional to a $6 \times 6$ unit matrix.

One observes that  if $X^f_{L,R}$ are set to be unit matrix, there exists only FCNC interaction of $a$ with neutrinos but no interaction with quarks and charged leptons, because
$V^\nu_{LL, RR, LR, RL}$ are separately not unitary.
Working in the basis where $M_e$ and $M_R$ are diagonalized, one can approximate~\cite{Abada:2007ux,He:2009tf} $V_{LL} = (1-\epsilon/2)V_{PMNS}$ with $\epsilon = Y_DM_R^{-2} Y^\dagger_D v^2/2$. Global fit finds that the matrix elements in $\epsilon$ are ${\cal O}(10^{-3})$~\cite{Fernandez-Martinez:2016lgt}. Therefore, the couplings $V^\nu_{LL}{V^\nu_{LL}}^\dagger$ are allowed at the level of $10^{-3}$. If different singlets are introduced for corresponding right-handed neutrinos to have different lepton numbers, one would need to change the Majoron couplings to light neutrinos to $V^\nu_{LL}X^\nu_R{V^\nu_{LL}}^\dagger$ with $X^\nu_R$ a diagonal matrix but different diagonal entries.  The individual off-diagonal couplings can be much larger than $10^{-3}$.
In general, the off-diagonal entries are arbitrary and should therefore be constrained by data. There are also constraints from mixing between heavy and light neutrinos.  However, they can be independent from light neutrino mixings and need to be constrained using data~\cite{He:2009ua}.

Before closing this section, we would like to make a comment about theories with spontaneous CP violation and how to identify the GB in the model. Spontaneous CP violation requires more than one Higgs doublet. When a global $U(1)_G$ is imposed, there may need more Higgs bosons to construct a model consistent with data~\cite{Geng:1988ty,He:1988dm}.
 For the model in Ref.~\cite{Geng:1988ty}, it is based on two Higgs doublet fields $H_j$ and two scalar singlet fields $S_j$ to incorporate spontaneous CP violation and PQ mechanism with invisible axion. The model in Ref.~\cite{He:1988dm} achieves spontaneous CP violation by adding another doublet and introducing one singlet. But in both models  each type of  the three generations of fermions has the same PQ charge, therefore  it does not have FCNC GB interaction. However, in this case the vevs are complex, that is, $v^q_{jk}$ becomes $v^q_{jk} e^{i\theta^q_{jk}}$. This may be more complicated in identifying the physical GB.

In this case the $z$ and $A$ become in the basis $-i (h_{jk}^q + i I^q_{jk})$,
\begin{eqnarray}
&&\vec z = (v^u_{jk}e^{i\theta^u_{jk}},\; v^d_{jk}e^{i\theta^d_{jk}},\; v^e_{jk}e^{i\theta^e_{jk}},\; v^\nu_{jk}e^{i\theta^\nu_{jk}},\; 0)\;,\nonumber\\
&&\vec A = (-(X^{u j}_L - X^{u k}_R)v^u_{jk}e^{i\theta^u_{jk}},\; (X^{d j}_L - X^{d k}_R)v^d_{jk}e^{i\theta^d_{jk}},\; (X^{e j}_L - X^{ e k}_R)v^e_{jk}e^{i\theta^e_{jk}} ,\;\nonumber\\
&&\hspace{0.7cm}\;-(X^{\nu j}_L - X^{\nu k}_R )v^\nu_{jk}e^{i\theta^\nu_{jk}},\; -(X^{\nu j}_R +X^{\nu k}_R)v^s_{jk}e^{i\theta^s_{jk}})\;.
\end{eqnarray}
The physical GB field is now
\begin{eqnarray}
a = {1\over N_\alpha}  Im \left [ \left ((X^{p l}_L - X^{p m}_R)   -  (X^{qj}_{L} - X^{qk}_R)\right ) (v^p_{lm})^2 v^{q}_{jk}e^{i\theta^q_{jk}}sign(q)(h^q_{ij}+iI^q_{ij}) + (v^p_{lm} )^2(X^{\nu j}_R +X^{\nu k}_R) v^s_{jk} e^{\theta^s_{jk}}(h^s_{jk} + iI^s_{jk})\right ]\;. \nonumber\\
\end{eqnarray}

This leads to the same $j_{af}^\mu$ as discussed before. We therefore conclude that no new CP violation phases for GB interactions with fermions arise. Some special cases of this type of models have been discussed in Ref.~\cite{Chen:2007nx,He:2010hz,Pan:2020qqd}. Three Higgs doublets $H_j$ and one complex scalar singlet S are introduced by setting $Q_{\mathrm{L}}(0)$, $U_{\mathrm{R}}(\pm 1)$, $D_{\mathrm{R}}(\pm 1)$, $L_L(0)$, $E_R(\pm 1)$, $H_{1,2}(+1)$, $H_{3}(-1)$, $S(+2)$ to get their model. The resulting
$j_{af}$ are special cases of Eq.~(\ref{mast-eq}) with same $U(1)_G$ charge for each type of  the three generations of fermions, therefore there are no FCNC GB interactions in the models.

\section{Special features for seesaw and Left-Right symmetric models}

We now discuss some interesting features of flavor changing GB interactions with fermions in some of the popular models, the Type-II, -III seesaw, and Left-Right symmetric models.

\subsection{Type-II seesaw model}

The simplest realization of Type-II seesaw~\cite{Schechter:1981cv,Magg:1980ut,Cheng:1980qt,Mohapatra:1980yp,Lazarides:1980nt} is by introducing a triplet Higgs field $\chi:(1,3,1)(-2)$ that couples to the neutrinos to give neutrino mass when $\chi$ develops a vev $v_{\chi}/\sqrt 2$ via the term $\bar L^c_L \chi L_L$. There is no need of introducing right-handed neutrino $\nu_R$ as in Type-I seesaw model. To have a GB, the Majoron in this case, one can impose the global lepton number conservation in the potential~\cite{Gelmini:1980re,Georgi:1981pg}. Since the $\chi$ field has a non-zero lepton number, its vev breaks both electroweak symmetry and global lepton number. The Goldstone boson ``eaten'' by Z boson is given by $z = (v I + 2 v_\chi I_\chi)/\sqrt{v^2+4 v^2_\chi}$. The Majoron is the another orthogonal component $(2v I -  v_\chi I_\chi)/\sqrt{v^2+4 v^2_\chi}$ whose coupling to neutrinos is proportional to the neutrino mass matrix. The mixing will induce Majoron to couple to charged leptons and quarks. Since the vev of $\chi$ is constrained to be less than a few GeV from the precise measurement of $\rho$ parameter~\cite{Zyla:2020zbs}, therefore the couplings of GB to charged leptons and quarks are small. There are no FCNC GB interactions. To remedy the problems related to light degrees of freedom in the model, one can introduce a singlet $S$ of the type discussed in Type-I seesaw model which couples to $\chi$ and $H$ through the term $H\chi H S$. But this still will not induce FCNC GB interactions.

If $L_j$ have different $U(1)_G$ as discussed in the general GB model in section II, there is the need to introduce several $\chi$ fields with the $U(1)_G$ charges $X_\chi^{jk} = - (X^{\nu j}_L + X^{\nu k}_L)$ and also to extend $S$ to $S^{jk}$. The term $H\chi H S$ is changed to $H_j\chi_{lm} H_k S^{pq}$ with the indices contracted in all possible ways for SM gauge group and also $U(1)_G$ singlets. In this case, following procedures in section II, we obtain the GB-neutrino current
\begin{eqnarray}
 j^\mu_{a\nu} =  {\bar v^2\over N_\alpha} \bar \nu_L \gamma^\mu \nu_L + {v^2\over N_\alpha} \bar \nu_LX^\nu_L \gamma^\mu \nu_L \;.
 \end{eqnarray}
Here
$v^2 = (v^u_{jk})^2 + (v^d_{jk})^2 +(v^e_{jk})^2 +4 (v^\chi_{jk})^2$ and
$\bar v^2 = (-(X^{uj}_L-X^{uk}_R) (v^u_{jk})^2 + (X^{dj}_L-X^{dk}_R) (v^d_{jk})^2 + (X^{ej}_L-X^{ek}_R) (v^e_{jk})^2
-2 (X^{\nu j}_L + X^{\nu k}_L) (v^\chi_{jk})^2$.

If $X_L^\nu$ is not  proportional to unit matrix, FCNC interactions will emerge. In the neutrino mass eigen-state basis, we have
\begin{eqnarray}
 j^\mu_{a\nu} =  {\bar v^2\over N_\alpha} \bar \nu_L \gamma^\mu \nu_L + {v^2\over N_\alpha} \bar \nu_L V_{PMNS}X^\nu_L \gamma^\mu V^\dagger_{PMNS} \nu_L \;,
 \end{eqnarray}
where $V_{PMNS}$ is the lepton mixing matrix.

At least two triplet fields $\chi$ with different $U(1)_G$ charges need to be introduced to have FCNC interaction. If the quark  and charged lepton $U(1)_G$ charges are also similarly the general model discussed, their corresponding couplings to the GB will be given by Eq.~(\ref{jac}) which lead to FCNC GB interaction with fermions in general.

\subsection{Type-III seesaw model}

In Type-III seesaw model~\cite{Foot:1988aq}, one replaces the right handed neutrinos $\nu_R$ by the $SU(2)_L$ triplet $\Sigma_L^c =\Sigma_R$, the charge conjugation of $\Sigma_L$,  transforming as a $(1,3,0)$ under the SM gauge group. It carries a $U(1)_G$ charge $X_R^\nu$ as in the Type-I seesaw model. The component fields are as the following
\begin{eqnarray}
\Sigma_L = \left ( \begin{array} {cc}
\Sigma_L^0/\sqrt{2}&\;\;\Sigma^+_L\\
\Sigma^-_L&\;\;-\Sigma^0_L/\sqrt{2}
\end{array}
\right )\;,\;\;\;\;
\Sigma_R = \left (\begin{array} {cc}
\Sigma_L^{0\; c}/\sqrt{2}&\;\;\Sigma^{-\; c}_L\\
\Sigma^{+ \; c}_L&\;\;-\Sigma^{0\;c}_L/\sqrt{2}
\end{array}
\right )\;.
\end{eqnarray}
We will rename them with
$\nu_R = \Sigma_L^{0\; c}$, $\psi_L = \Sigma^-_L$ and $\psi_R = \Sigma^{+\;c}_L$.

The Yukawa interaction terms are given by
\begin{eqnarray}
L = - \bar Q_L^j Y^{jk}_u \tilde H^u_{jk} U_R^k - \bar Q_L^j Y^{jk}_d H^d_{jk} D_R^k - \bar L_L^j Y^{jk}_e H^e_{jk} E_R^k
- \bar L_L^j \sqrt{2} Y^{jk}_\nu \Sigma_R^k \tilde H^\nu_{jk} - {1\over 2} Tr \bar \Sigma_R^{jc} Y^{jk}_s S_{jk} \Sigma^k_R+ \mbox{H.c.}\;.
\end{eqnarray}
The GB field is in general given by Eq.~(\ref{axion-field}). The GB couplings to up- and down-type quarks and also to neutrinos are the same as those given in Type-I seesaw model. But the couplings to charged leptons will be modified because of the existence of $\psi_{L,R}$. We have the mass and GB interaction terms
\begin{eqnarray}
L =&& -(\bar E_L, \bar \psi_L) M_c \left ( \begin{array}{c}
E_R\\
\psi_R
\end{array}
\right )\nonumber\\
&&-i a (\bar E_L, \bar \psi_L) \left ( \begin{array}{cc}
M_e {\bar v^2\over N_\alpha} - {v^2\over N_\alpha} (X_L^e M_e - M_e X^e_R)&\;\; \sqrt{2} M_D{\bar v^2\over N_\alpha}
-{v^2\over N_\alpha} (X^e_L \sqrt{2}M_D - \sqrt{2} M_D X^\nu_R)\\
\\
0&\;\;{v^2\over N_\alpha}( X^\nu_R M_R + M_R X^\nu_R)
\end{array}
\right ) \left ( \begin{array}{c}
E_R\\
\psi_R
\end{array}
\right ) +\mbox{H.c.}\;.
\end{eqnarray}
where
\begin{eqnarray}
M_c = \left ( \begin{array}{cc}
M_e&\;\; \sqrt{2} M_D\\
0&\;\;M_R
\end{array}
\right )\;.
\end{eqnarray}

Using the equations of motion, the GB current $j^\mu_e$  in the  interaction $\partial_\mu a\; j^\mu_e$, can be written as
\begin{eqnarray}
j^\mu_e &=& - {\bar v^2\over   N_\alpha} \bar E_L\gamma^\mu E_L + {v^2\over N_\alpha} (\bar E_L X^e_L \gamma^\mu E_L + \bar E_R  X^e_R\gamma^\mu E_R)
+{v^2\over N_\alpha}(\bar \psi_R X^\nu_R \gamma^\mu \psi_R -\bar \psi_L  X^\nu_R \gamma^\mu \psi_L )\;.
\end{eqnarray}
One can easily see that GB will have FCNC interactions with charged leptons too.

We would like to mention a special feature noticed recently in Ref.~\cite{Cheng:2020rla} which can be achieved by just introducing one $S$ to the usual Type-III seesaw model, and normalizing  $f_a$ to be equal to $v_s$ as that in Ref.~\cite{Cheng:2020rla} by choosing $X^e_{L,R}=X^\nu_R=1/2$. In this case $\bar v^2=0$.
Using vector current conservation $\partial_\mu (\bar E\gamma^\mu E + \bar \psi \gamma^\mu \psi) = 0$, we have
\begin{eqnarray}
j^\mu_{e} = -{v^2\over N_\alpha}\bar \psi_L \gamma^\mu \psi_L\;.
\end{eqnarray}
The mass matrix $M_c$ can be diagonalized in the form $M_c = {V^{e\,L}}^\dagger \hat M_c V^{e\,R}$. Here $V^{e \, L(R)}$ are $6\times 6$ unitary matrices.
Writing $V^e$ into blocks of $3\times 3$ matrices, we have
\begin{eqnarray}
V^{e\;L(R)} = \left( \begin{array}{ll}
V^{e\;L(R)}_{LL}\;\;&V^{e\;L(R)}_{LR}\\
\\
V^{e\;L(R)}_{RL}\;\;&V^{e\;L(R)}_{RR}
\end{array}
\right) .
\end{eqnarray}
We then obtain Majoron $J$ interactions with neutrinos and charged leptons in the mass basis as
\begin{eqnarray}
{\partial_\mu J\over 2 f_J}
\left [
 - 2  ( \bar E_L \gamma^\mu V^{e\;L}_{LR}{V^{e\;L}_{LR} }^\dagger E_L + \bar \psi_L  \gamma^\mu V^{e\;L}_{RR} {V^{e\;L}_{LR}}^\dagger E_L
 + \bar E_L \gamma^\mu V^{e\;L}_{LR} {V^{e\;L}_{RR}}^\dagger \psi_L + \bar \psi_L \gamma^\mu V^{e\;L}_{RR}  {V^{e\;L}_{RR}}^\dagger \psi_L ) \right ].
\end{eqnarray}
 The size of off-diagonal entries is as large as the level of  $10^{-3}/f_J$, similar to that in Type-I seesaw model. If there are more than one singlet with different lepton numbers and different right-handed neutrinos are assigned with different lepton numbers, one would need to change the Majoron couplings to light neutrinos to $V^\nu_{LL}X^\nu_R{V^\nu_{LL}}^\dagger$ with $X^\nu_R$ a diagonal matrix but different diagonal entries.  The individual off-diagonal couplings can be much larger than $10^{-3}/f_J$. In this model, the GB is a typical Majoron whose FCNC interactions with fermions can lead to interesting consequences as shown in Ref.\cite{Cheng:2020rla}.

 We note in passing that because of the appearance of new particle $\psi$ in the theory, the GB-two-photon coupling in Type-III seesaw model will be modified compared with that in Type-I seesaw model. One needs to add a new term  $\frac{v^2}{N_\alpha} Tr( X^\nu_R) Q^2_\psi N_c^\psi$ into $\tilde E(X)$ for $a F_{\mu\nu}\tilde F^{\mu\nu}$ in Eq.~(\ref{Lagamma}).

\subsection{Left-Right symmetric model}

For Left-Right symmetric model, the gauge group is extended from the SM gauge group to $SU(3)_C \times SU(2)_L \times SU(2)_R\times U(1)_{B-L}$~\cite{Mohapatra:1980yp,Mohapatra:1974gc,Senjanovic:1975rk}. The left-handed quarks $Q_L$ and leptons $L_L$ transform as $(3,2,1,1/6)$ and $(1,2,1,-1/2)$. The right-handed quarks $Q_R$ and leptons $L_R$ are grouped into doublets of $SU(2)_R$, and transform as $(3,1,2,1/6)$ and $(1,1,2,-1/2)$. If a global $U(1)_G$ imposed on the model is broken, a GB will arise. We will indicate the $U(1)_G$ charges similarly as what we have done in section II.

To have a GB symmetry in the Left-Right symmetric model, at least two bi-doublets $\phi_{1,2}$ transforming as $(1,2,2,0)$ with different $U(1)_G$ charges need to be introduced in order to have phenomenologically acceptable quark mass matrices and mixing. This also implies different generations of quarks and also leptons, some of them, should have different $U(1)_G$ charges. We will construct a minimal
model which also has triplets $\Delta_L: (1,3,1,1)$ and $\Delta_R: (1,1,3,1)$ to make effective the seesaw mechanism. It turns out at least two different sets of triplets are needed to make the resulting $U(1)_G$ invisible as in the sense of DFSZ type~\cite{Grimus:1982qu}.

As an example, the $U(1)_G$ charges for various fermions and scalars as well as their Left-Right components can be set as below
\begin{eqnarray}
&&Q_{L1}: (0),\;\;Q_{L 2,3}: (X),\;\;Q_{R1}: (0),\;\;Q_{R2,3}: (-X),\;\;L_{L1}: (0),\;\;L_{L 2,3}: (X),\;\;L_{R1}: (0),\;\;L_{R2,3}: (-X),\nonumber\\
&&\phi_1: (X),\;\;\phi_2: (2X)\;,\;\;\Delta_{L1}: (X),\;\;\Delta_{L2}: (2X),\;\;\Delta_{R1}: (-X),\;\;\Delta_{R2}: (-2X).
\end{eqnarray}
 We take this type of model  as an example to work out some details. With different assignment of $U(1)_G$ charges for fermions, the resulting Yukawa texture will be different. 
Therefore this illustrates how to construct a realistic Left-Right symmetric model with an additional global $U(1)_G$ symmetry broken spontaneously.

We write the bi-doublets as: $\phi_{1,2} = \left ( \tilde \phi_{1,2}, \bar \phi_{1,2} \right )$,
where $\tilde \phi_j = i\sigma_2 \phi^*_j$. Both $\phi_j$ and $\bar \phi_j$ are doublets of $SU(2)_L$. Writing in this way enables us to use directly the results obtained before for finding GB field since they both transform the same under $SU(2)_L$. The components of these fields are
\begin{eqnarray}
&&\phi_j = \left ( \begin{array}{c}
h^+_j\\
\\
{v_j\over \sqrt{2}}(1+ {h_j\over v_j} + i {I_j\over v_j})
\end{array} \right ),\;\;
\bar \phi_j = \left ( \begin{array}{c}
\bar h^+_j\\
\\
{\bar v_j\over \sqrt{2}}(1+ {\bar h_j\over \bar v_j }+ i {\bar I_j\over \bar v_j})
\end{array} \right ),\nonumber\\
&&\nonumber\\
&&\Delta_{Lj} = \left ( \begin{array}{cc}
{1\over \sqrt{2}}\delta^+_{Lj}&\;\;\delta^{++}_{Lj}\\
\\
{v_{Lj}\over \sqrt{2}}(1+ {\delta^0_{Lj}\over v_{Lj}} + i {I_{Lj}\over v_{Lj}})&\;\;- {1\over \sqrt{2}}\delta^+_{Lj}
\end{array} \right ),\;\;
\Delta_{Rj} = \left ( \begin{array}{cc}
{1\over \sqrt{2}}\delta^+_{Rj}&\;\;\delta^{++}_{Rj}\\
\\
{v_{Rj}\over \sqrt{2}}(1+ {\delta^0_{Rj}\over v_{Rj}} + i {I_{Rj}\over v_{Rj}})&\;\;-{1\over \sqrt{2}}\delta^+_{Rj}
\end{array} \right ).
\end{eqnarray}

The Yukawa interactions are given by
\begin{eqnarray}
L_Y =& -& \bar Q_L (\kappa^q_1 \phi_1 + \kappa^q_2 \phi_2) Q_R - \bar L_L ( \kappa^l_1 \phi_1 + \kappa_2^l \phi_2) L_R
\nonumber\\
&-& \bar L_L^c (Y_{L1} \Delta_{L1} + Y_{L2} \Delta_{L2}) L_L - \bar L_R^c (Y_{R1} \Delta_{R1} + Y_{R2} \Delta_{R2}) L_R+\mbox{H.c.}\;.
\end{eqnarray}
If there is just one bi-doublet, only one of the $\kappa$ terms is allowed for the quark and lepton sectors because of the non-zero $U(1)_G$ charges. This leads to the up and down sector of quark mass matrices to be proportional each other, which results in unrealistic mass relations without mixing. This is the reason that one needs to have more than one bi-doublet.
Because of the $U(1)_G$ charges assigned, the $\kappa$ and $Y$ have the following forms
\begin{eqnarray}
&&\kappa^{q,l}_1 = \left ( \begin{array}{ccc}
0&\;\;K^{q,l}_{12}&\;\;K^{q,l}_{13}\\
K^{q,l}_{21}&\;\;0&\;\;0\\
K^{q,l}_{31}&\;\;0&\;\;0
\end{array} \right ),
\;\;\;\;\;\;\;\;\;\kappa^{q,l}_2 = \left ( \begin{array}{ccc}
0&\;\;0&\;\;0\\
0&\;\;K^{q,l}_{22}&\;\;K^{q,l}_{23}\\
0&\;\;K^{q,l}_{32}&\;\;K^{q,l}_{33}
\end{array} \right ),\nonumber\\
&&Y_1^{L,R} = \left ( \begin{array}{ccc}
0&\;\;Y^{L,R}_{12}&\;\;Y^{L,R}_{13}\\
Y^{L,R}_{12}&\;\;0&\;\;0\\
Y^{L,R}_{13}&\;\;0&\;\;0
\end{array} \right ),
\;\;Y^{L,R}_2 = \left ( \begin{array}{ccc}
0&\;\;0&\;\;0\\
0&\;\;Y^{L,R}_{22}&\;\;Y^{L,R}_{23}\\
0&\;\;Y^{L,R}_{23}&\;\;Y^{L,R}_{33}
\end{array} \right ).
\end{eqnarray}

We will assume $v_{Lj} = 0$, the quark mass matrices $M_{u,d}$ and the lepton mass matrices $M_e$ and $M_\nu$ are given by
\begin{eqnarray}
&&M_u = {\kappa^q_1 v_1\over \sqrt{2}} + {\kappa^q_2 v_2\over \sqrt{2}},\;\;M_d = {\kappa^q_1 \bar v_1\over \sqrt{2}} +{\kappa^q_2 \bar v_2\over \sqrt{2}}\;;\;\;M_e = {\kappa^l_1 \bar v_1\over \sqrt{2}} +{\kappa^l_2 \bar v_2\over \sqrt{2}},\nonumber\\
&&\nonumber\\
&&M_\nu = \left ( \begin{array}{cc}
0&\;\; M_D\\
M^T_D& M_R
\end{array} \right ),\;\;
\mbox{with}\;\;
M_D = {\kappa^l_1 v_1\over \sqrt{2}} + {\kappa^l_2 v_2\over \sqrt{2}},\;\;
M_R = {Y_{R1} v_{R1}\over \sqrt{2}} + {Y_{R2} v_{R2}\over \sqrt{2}}.
\end{eqnarray}

We now work out the GB fields following the method previously used. The vevs of $\Delta_{Ri}$ break $SU(2)_R$ and also $U(1)_{B-L}$, and the vevs of $\phi_{1,2}$ break both the $SU(2)_R$ and $SU(2)_L$, and all of them also break $U(1)_G$. For working out the physical GB, we choose three broken generators $I^L_3$, $B-L$ and $A$ of $I^L_3$, $B-L$ and $U(1)_G$ symmetries as
\begin{eqnarray}
I^L_3 : (v_1, \;\bar v_1, \;v_2,\;\bar v_2, 0,\;0),\;\;B-L : (0,\;0,\;0,\;0,\;v_{R1}, \;v_{R2}),\;\;A : (-v_1, \;\bar v_1,\;-v_2,\;\bar v_2, v_{R1}, \;2v_{R2})\;.
\end{eqnarray}
The physical GB will be the linear combination with its orthogonal to $I^L_3$ and $B-L$. We have
\begin{eqnarray}
a:  \left (-v^2_R \bar v^2 I^L_3 + v^2 \bar v^2_R (B-L) - v^2 v^2_R A\right ),
\end{eqnarray}
where $v^2 = v^2_1+\bar v^2_1 + v^2_2 + \bar v^2_2$, $\bar v^2 = v^2_1-\bar v^2_1 + v^2_2 - \bar v^2_2$ and $v_R^2 = v^2_{R1} + v^2_{R2}$ and $\bar v^2_R = v^2_{R1} +2 v^2_{R2}$.
Expressing $a$ in terms of $I_j$ field of the various scalars, we have
\begin{eqnarray}
a = {1\over N_\alpha} &&\left [ -v^2_R (\bar v^2 - v^2)v_1 I_1  -v^2_R (\bar v^2 -  v^2)v_2 I_2
- v^2_R (\bar v^2 + v^2) \bar v_1 \bar I_1  - v^2_R (\bar v^2 +  v^2) \bar v_2 \bar I_2 \right .\nonumber\\
&&+ \left .v^2(\bar v^2_R - v^2_R) v_{R1}I_{R1} + v^2(\bar v^2_R - 2v^2_R) v_{R2}I_{R2 }\right ].
\end{eqnarray}
Note that if there is only one $\Delta_{Rj}$ or both of $\Delta_{R_j}$ have the same $U(1)_G$ charge, there is no $I_{Rj}$ in $a$, then the axion decay constant is order $v$ which is a visible axion type.

We obtain the GB currents for charged fermions and neutrinos in the form given  in Eqs.~(\ref{jac}) and (\ref{janu}) with
\begin{eqnarray}
X^u_L =X^d_L= X^e_L=X^\nu_L=\left (\begin{array}{ccc}
0&\;0&\;0\\
0&\;-1&\;0\\
0&\;0&\;-1
\end{array}
\right ),\;\;X^u_R =X^d_R=X^e_R= X^\nu_R=\left (\begin{array}{ccc}
0&\;\;0&\;\;0\\
0&\;\;1&\;\;0\\
0&\;\;0&\;\;1
\end{array}
\right ),
\end{eqnarray}
and for $u$, $d$ and $e$ replace $\bar v^2/N_\alpha$ and $v^2/N_\alpha$ by $-v^2_R \bar v^2/N_\alpha
$ and$-v^2_Rv^2/N_\alpha$. Also for right handed neutrinos, replace $(v^2/N_\alpha) X^\nu_R$ by $v^2(\bar v^2_R- v^2_R)  X^\nu_R/N_\alpha$.

\section{Discussions and Conclusions}

We have carried out a systematic model building study for  FCNC GB interactions in the three generations of fermion sectors, or separately in the quark, charged lepton and neutrino sectors. It is based on renormalizable models with an additional $U(1)_G$ global symmetry which is spontaneously broken besides the gauge symmetries of the model. Several popular models have been discussed.

To study how FCNC GB interactions emerge, we have developed a method to identify the GB in a beyond  SM with an additional $U(1)_G$ global symmetry which is broken by an arbitrary number of Higgs bosons. Although our main aim is to study how FCNC GB interactions emerge, we find that our method can be used easily to build a desired model and to provide some insight about some general properties of GB interactions in a simple fashion.
Many models studied in the literature can be easily reproduced by just assigning the appropriate $U(1)_G$ charges as discussed in the previous sections. We  also provide a general proof of the equivalence of using physical GB components and GB broken generators for calculating GB couplings to two gluons and two photons, although they have different form. The final results only depend on the $U(1)_G$ charges $X^f_{L,R}$ and the kind of colored, and charged particles in the model. Parameters in the FCNC interactions do not affect GB interactions with two gluons and two photons. We have shown that for spontaneous CP violation models, there is no new CP violating phase of GB-fermions interactions.

For FCNC GB interactions with fermions, we find that there are two types of sources. One of them is that different generations of fermions have different $U(1)_G$ charges, and another is due to mass splits of left- and right-handed particles, like
neutrino masses in Type-I and Type-III seesaw models.
Even if all generations have the same $U(1)_G$ charges, there still are in general FCNC GB interactions  with neutrinos which have not been studied carefully previously.
For Type-III seesaw model, there are also FCNC GB interactions with charged leptons.
For Type-II seesaw model, at least two triplets are needed to have FCNC GB interactions with fermions.
For Left-Right symmetry model, to make FCNC GB interactions with fermions to be invisible, at least two bi-doublets plus more than one triplets scalars need to be introduced similar as that in DFSZ model.

Whether or not fundamental GB exist is of course an experimental issue. Several high luminosity facilities in running, such as the BESIII, LHCb, BELLE-II, will provide us with more information. We eagerly wait for more data to come to test models having FCNC GB interactions with fermions.

\acknowledgements

This work was supported in part by NSFC (Grants 11735010, 11975149, 12090064), by Key
Laboratory for Particle Physics, Astrophysics and Cosmology, Ministry of Education, and Shanghai
Key Laboratory for Particle Physics and Cosmology (Grant No. 15DZ2272100), and in part
by the MOST (Grants No.109-2112-M-002-017-MY3).


\end{document}